**CryptoHalal: An Intelligent Decision-System for Identifying Halal and Haram Cryptocurrencies**

Shahad Z. Al-Khalifa[*]

Computer Science, College of Computer and Information Sciences

King Saud University

November 4, 2022

---

[*] Email address: shahadalkhalifa90@gmail.com
Acknowledgments: I would like to thank my supervisor, Professor Hend Al-Khalifa, for her guidance through each stage of the process, and whose expertise was valuable in formulating the research.

# Abstract

In this research, we discussed a rising issue for Muslims in today's world that involves a financial and technical innovation, namely: cryptocurrencies. We found out through a questionnaire that many Muslims are having a hard time finding the jurisprudence rulings on certain cryptocurrencies. Therefore, the objective of this research is to investigate and identify features that play a part in determining the jurisprudence rulings on cryptocurrencies. We have collected a dataset containing 106 cryptocurrencies classified into 56 Halal and 50 Haram cryptocurrencies, and used 20 handcrafted features. Moreover, based on these identified features, we designed an intelligent system that contains a Machine Learning model for classifying cryptocurrencies into Halal and Haram.

# Table of Contents

# List of Tables

# List of Figures

**Introduction**

Technology advancement has brought rapid changes in human activities introducing innovations in various fields. The financial sector has benefited greatly from the digitalized services offered by recent digital transformation. *Cryptocurrency* is one of the recent innovations that has taken the world of finance by storm, considering its free-flowing trading system without banking fees and users can exchange cryptocurrency digitally without a third-party oversight (Al-hussaini et al., 2019).

Due to the rise of cryptocurrency users and traders in the past decade, thousands of cryptocurrencies have been created, and new currencies enter the field every day. According to *CoinMarketCap*[1], the world's most-referenced price-tracking website for crypto assets, more than 30 new cryptocurrencies are added to the website every month.

The first cryptocurrency, namely Bitcoin, was created by Satoshi Nakamoto in 2009 (Billah, 2019). Figure 1 shows the rise of Bitcoin trading volume in the period spanning 2009 through 2022. The volume indicates how many Bitcoins are sold and bought on specific exchanges. The figure also shows that the volume began to get higher in the last five years, stating the beginning of a new era of financial technology. Trading volume reached its highest with a massive peak at the end of the year 2017 to early 2018. Many sources have stated that it was a growing interest from professional investors as an opportunity for a long-term investment.

---

[1] https://coinmarketcap.com/ [Date accessed: 19/01/2022]

**Figure 1**

*The total USD value of trading volume on major bitcoin exchanges*

*[Source: https://blockchain.com/charts/trade-volume]*

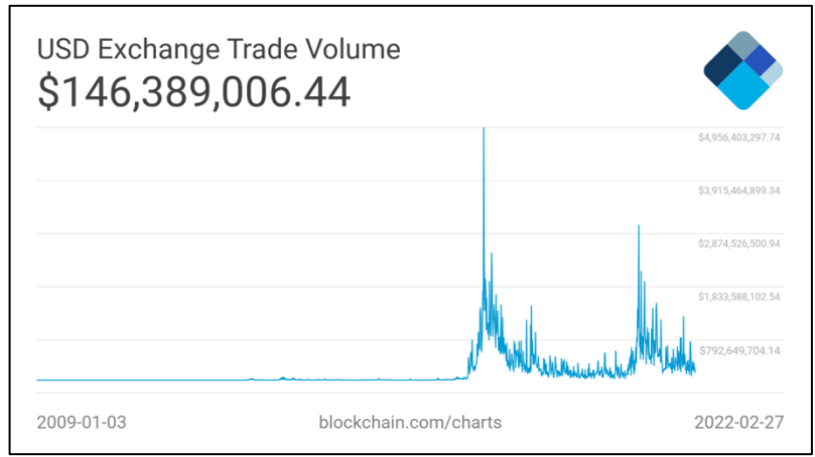

From the Islamic finance perspective, cryptocurrencies have raised many issues among Islamic scholars regarding its compatibility with *Sharia* (شَرِيعَة), particularly if it satisfies the Islamic requirement for a currency such as intrinsic value. Recently, the Muslim community has shed light on the matter of cryptocurrency. In 2021, two important events took place in Saudi Arabia (SA), where financial topics were discussed, and among them was Sharia's view on cryptocurrencies. The first event, the Cryptocurrency Seminar, took place in Jeddah, SA organized by the International Islamic Fiqh Academy (IIFA)[2]. The seminar's main goal was to discuss the concept of cryptocurrencies, their ruling statements, and the economic and Sharia perspective on the future of cryptocurrencies.

The second event, the 15th Islamic Financial Services Board (IFSB) summit hosted by the Saudi Central Bank (SAMA) in Jeddah, SA, under the theme "Digital Transformation of Islamic Finance: Innovation and Resilience"[3]. The summit discussed several topics, including the implication of crypto assets in Islamic finance and digital financial transformation.

---

[2] (ندوة العملات الرقمية المشفّرة) https://iifa-aifi.org/ar/28224.html [Date accessed: 19/2/2022]
[3] https://www.ifsbksa2021.com/page/about [Date accessed: 8/2/2022]

It has become necessary for Islamic finance to adapt to modern changes and assist in providing products and services to Muslims that are compliant with Sharia. Hence, the objective of this research is to investigate and identify features and characteristics that play a part in determining the jurisprudence rulings on cryptocurrencies. Furthermore, we focus on designing a platform that employs a trained machine learning model to predict whether a cryptocurrency is *Halal* or *Haram* based on its project and features that will be further explained in the upcoming sections.

The idea of this research was inspired by people's recurring need for a platform to answer their questions about the compliance of specific cryptocurrencies with Sharia. To justify the need for this type of research, we conducted a short questionnaire spread among Muslim cryptocurrency investors on Twitter. The questionnaire received a total of 748 responses. Around 96% of the respondents have cryptocurrencies, and nearly half of them own more than 5 cryptocurrencies, as shown in Figure 2. We asked a question to see if our respondents are interested in the jurisprudence ruling on their own cryptocurrencies. Figure 3 indicates that most of the respondents are interested in knowing the ruling of their owned cryptocurrencies. The next question asked about their way of searching for the ruling of the cryptocurrencies. Approximately 64% of the respondents answered that they searched about the cryptocurrency project themselves on the internet and read about it for any prohibited activities. In contrast, the rest of the respondents' answers were distributed among Telegram channels and Twitter, directly asking Islamic scholars, and asking friends and family. Lastly, we asked them about their opinion regarding having an automated system for extracting the jurisprudence ruling of cryptocurrencies. Remarkably, 86% agreed with this idea, while 12.7% were neutral, and only 4% of the respondents disagreed (See Figure 4). Most respondents agreed due to the left of burden in searching about cryptocurrencies' jurisprudence rulings to a trusted entity, on the other hand, the disagreement of respondents was due to their concern of trusting an automated system to conclude a jurisprudence ruling on cryptocurrencies.

**Figure 2**

*Responses to "How many cryptocurrencies do you own?"*

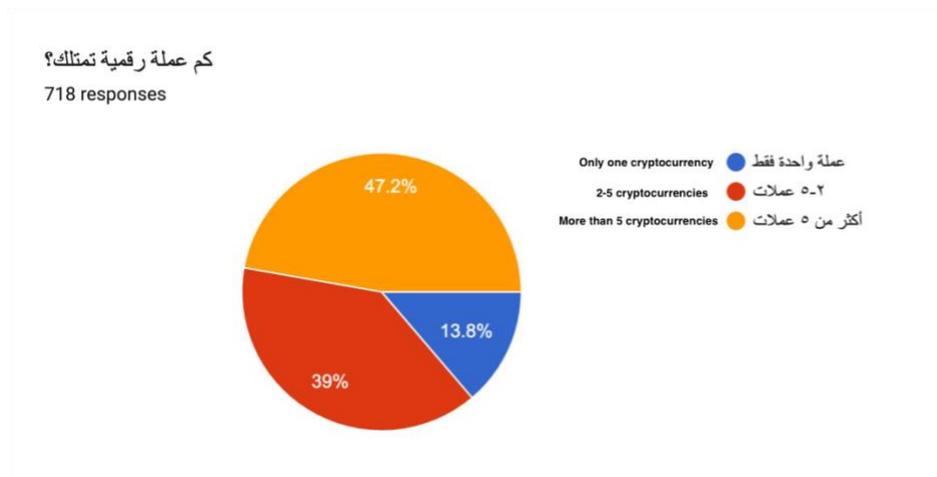

**Figure 3**

*Responses to "Are you interested in knowing the jurisprudence ruling for the cryptocurrency you own?"*

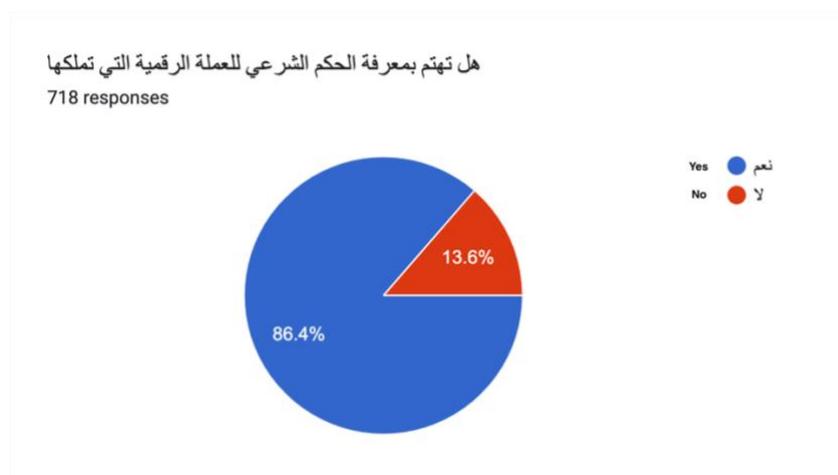

**Figure 4**

*Responses to "Do you think that having an intelligent platform that helps extract the jurisprudence ruling of cryptocurrencies would impact your decision to buy cryptocurrencies?".*

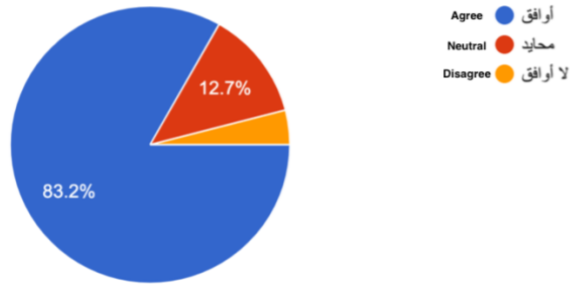

هل تظن بأن وجود منصة ذكية تساعد على استخراج الحكم الشرعي لعملة رقمية أوتوماتيكيا خلال ثواني سيُساعدك في قرار شراء العملة الرقمية؟

620 responses

Agree أوافق
Neutral محايد
Disagree لا أوافق

83.2%
12.7%

Since Fintech and cryptocurrencies play a vital role in the financial world today, there has been a greater demand for research approaches toward these topics from the Islamic finance perspective. Our questionnaire shows that such a system like ours is needed in the Muslim community and would solve a large-scale conflict in today's world. Thus, this research aims to explore the world of cryptocurrencies from the Islamic finance perspective. Moreover, the designed system represents a major step for Sharia in coping with digital financial development and advanced models based on Artificial Intelligence (AI) and Machine Learning. In addition, the published dataset that was used in this research may serve as a tool to be expanded and used for further research and discussion. This research will contribute by:

1. Identifying the Halal and Haram features of cryptocurrencies with the supported explanation.

2. Designing an intelligent system that will serve Muslims in identifying which cryptocurrencies are Halal or Haram to invest in.

3. Creating a dataset of Halal and Haram cryptocurrencies.

The remainder of this research is organized as follows. The second section discusses the background of cryptocurrencies and the technology behind them, the definition of money in Sharia's perspective, and the main pillars of money preservation in Islam. The third section is the literature review on Islamic scholars' views on cryptocurrency and the dynamics of the cryptocurrency market that threatens its credibility. The fourth section describes the collected data and the designed system with measured evaluation. The last section concludes the research with limitations and future directions.

**Background**

This section will provide a brief overview of the concept of cryptocurrency and the technology behind it and machine learning techniques. Then, we will discuss the types of digital currencies and some adapted protocols. Also, we will look into the Islamic finance perspective of cryptocurrencies as money and a medium of exchange in Sharia. Lastly, we will discuss the main pillars of money preservation in Islam and cryptocurrencies' place among them.

**Cryptocurrency Overview**

Cryptocurrency is an innovative digital currency created through cryptography and stored within a blockchain using encryption technology (Billah, 2019). The backbone of cryptocurrency is blockchain technology which is a distributed, autonomous, open-source software application and function that operates in a peer-to-peer network (Islam et al., 2018). Besides being a digital currency, cryptocurrencies are helping the financial technology system to innovate, transform and incorporate better understanding and accountability through the feature of the blockchain (Alzubaidi & Abdullah, 2017). The value of cryptocurrency is based on the blockchain creation to store the data, and the calculation is based on an algorithm. The mechanism of how the value of the cryptocurrency is created depends on how big the blockchain can create the data. The bigger it is, the higher the fee for miners (Siswantoro et al., 2020).

According to Asif (2018) and Wu et al. (2018), there are two types of digital currencies: coins and tokens. Coins operate on their independent network and are specifically created to serve as a medium of exchange. On the other hand, tokens operate on top of a coin network as a platform, and their value depends on the performance of the issued company. Thus, when investing in a token, it can be seen as an asset, much like owning a share in a company.

Bitcoin employs the *proof-of-work* (PoW) consensus model, which is a combination of cryptography and computational power that ensures the authenticity of the data recorded on the blockchain. For the model to prove that the block is valid, i.e., a transaction between two peers is successful, the nodes in the network called *miners* use computational powers to verify a transaction and compete to solve a cryptographic problem imposed by the PoW protocol. This process is called *mining* and it requires computers to run at a maximum capacity which consumes an enormous amount of time and energy which is proof that the work is done (Seang & Torre, 2018).

An alternative to PoW is *proof-of-stake* (PoS). In PoS, miners are not chosen according to their competence to solve a cryptographic problem, but according to how much stake they hold in the cryptocurrency. PoS is considered energy-efficient compared to PoW. However, PoS lacks an important feature that is provided in PoW protocols: *compounding*, which means that whenever a miner solves a cryptographic problem, the miner will earn a reward that will be added to the buyer's account which will increase the chance for the miner to be selected first in the future (Fanti et al., 2019).

**Machine Learning Techniques**

Machine Learning (ML) is considered a subset of AI that exhibits the experimental learning associated with human intelligence and can be defined as computational methods using provided data to improve performance or make accurate predictions. ML consists of designing accurate and efficient prediction learning algorithms that mainly rely on the used data to generate predictions on unseen data (Helm et al., 2020; Mohri et al., 2018). The learning algorithms can be categorized into four types, such as supervised, unsupervised, semi-supervised, and reinforcement learning (Sarker, 2021). In supervised learning, input data are labeled with a corresponding output and the goal is to construct an algorithm that maps inputs to outputs (Liakos et al., 2018). Supervised learning is considered a task-driven approach and the most common supervised tasks are *classification* that separates data, and *regression* that fits the data. On the other hand, unsupervised learning is considered a data-driven approach that uses unlabeled data to identify meaningful structures and trends in the data. Semi-supervised can be defined as a hybrid containing both supervised and unsupervised methods. Reinforcement learning is considered an environment-driven approach that enables machines to automatically evaluate the optimal behavior in a particular environment to improve its efficiency (Sarker, 2021).

**Money in Islam**

Throughout centuries, Islamic finance scholars have debated many topics, including financial matters, particularly money in all its forms. According to Uthmani (1998), Sharia does not identify money as a subject of trade as it has no intrinsic utility and is only a medium of exchange. Profit cannot be generated in exchange for one unit of money to another since each unit is equal to the other unit of the same denomination. Hence, profit is generated when something with intrinsic utility is sold for money. Ibn Taymiyah (ابن تيمية) supported that Sharia has not defined a specific condition or definition for currency or money, but instead left it to the understanding of the people[4].

From the Sharia perspective, a currency has three components: Property "*Mal*" (مال), legal value, and monetary value. As stated by Islam (1999), Mal is defined as a human tendency that can be stored over time, has been created for human beings' goodness, and is considered lawful in Islam. For example, alcohol is not considered Mal since it's prohibited in Islam. Therefore, the requirement of Mal to be exchanged needs to be legal or a subject for use in Sharia. As discussed previously regarding money being left to the understanding of people, Quran and Sunnah have not explicitly stated that there must be a tangible form for money to have value as it may be intangible in limitation for it to be desirable and retrievable.

Monetary value is the key element in an asset that is entitled to serve as money, and it has two primary functions, which are: an independent standard of value and a unit of account (Yuneline, 2019). Referring to Adam (2017), monetary value enables money to independently evaluate prices and rate goods. Since it's an independent standard value, it must have stability and should be universally acceptable. Whereas the unit of account refers to being the main reference point and benchmark for people to send prices and record debt.

---

[4] Ibn Taymiyyah, (2005) Majmu' al-Fatawa (مجموع الفتاوى), Egypt: Dar ibn Hazm.

**The Main Pillars of Money Preservation in Islam**

The main pillars of the preservation of money (حفظ المال) in Islam are Marketability and circulation, Transparency, Preservation, Durability, and Equity.

*Marketability and Circulation*

The entire notion of Sharia revolves around the preservation of order, prevention of prohibited activities, establishing equality among people, and obeying the law. Marketability promotes the circulation of wealth and prohibits the hoarding of speculation. Suppose a cryptocurrency is acquired for saving purposes. In that case, it might be prohibited in Sharia, as the high instability attached to the exchange rate involves uncertainty and risk "*Gharar*", as well as participating in speculation "*Maysir*" (Abubakar et al., 2018).

*Transparency*

The objective of transparency is to avoid harm and being a victim of manipulation. Since cryptocurrency transactions protect the anonymity of their participants and preserve their identity, this pillar might be violated.

*Preservation*

Preservation of money means protection of the wealth of the community from falling into the hands of intruders without compensation. Since the nature of cryptocurrencies is obscure, it is an attraction for hackers and frauds. In addition, if an economy prohibits the use of cryptocurrency, its price will crash, impacting the world's cryptocurrency market. This is seen as the primary reason why cryptocurrencies are uncertain and risky, which goes against the preservation of Mal.

*Durability*

Durability denotes that money should be earned legally in Sharia with avoidance of Gharar and debate. Some Islamic scholars viewed paper money, due to inflation, as fitting in this category, whereas Bitcoin and cryptocurrencies pass this condition. It is believed that cryptocurrencies hold less risk of forgery compared to paper money due to blockchain keeping the accurate amount of exchange in a transaction, as well as the slight to non-existent chance for cryptocurrencies to be forged if dealt with a trusted platform (Muedini, 2018).

*Equity*

Equity promotes the preservation of communal wealth by generating a just and fair system for all. However, Islamic scholars are questioning whether cryptocurrencies follow this category due to the anonymity and lack of transparency to people who are behind the scenes (Adam, 2017).

**Literature Review**

This section will review Islamic scholars' sources of jurisprudence rulings regarding cryptocurrencies. The views are composed of: proponents, opponents, and neutral (have not conducted an official ruling). Next, we will look into cryptocurrency credibility against scams and counterfeit cryptocurrencies that have caused massive financial losses among users. Cryptocurrency credibility has been included due to supporting material for the dynamics of the cryptocurrency market.

**Cryptocurrency Jurisprudence Ruling in Sharia**

Islamic scholars are still in debate on the jurisprudence ruling on cryptocurrencies. Since Bitcoin is the most famous cryptocurrency, the existing *Fatawa* فتاوى (Islamic scholars' legal opinions) are revolved around it. However, the views and arguments on any cryptocurrency are the same as Bitcoin, considering they serve the purpose of money and medium of exchange. Based on the literature, there are three views on the ruling of cryptocurrency: The first view states that it is compliant with Sharia to buy and trade with cryptocurrencies as long as the cryptocurrency project is permissible in Islam. The second view states that cryptocurrencies are speculative, similar to gambling, and thus are forbidden in Islam. The last view states that the matter of cryptocurrency is still not transparent to Islamic scholars and needs further study and investigation. Next, we will present the views of the proponent, opponent, and neutral sources.

### *Proponents*

Some Islamic scholars believe that cryptocurrencies are compliant with Sharia. For example, according to the modern Islamic scholar Alsulmi (2020), cryptocurrencies such as Bitcoin are a commodity and are used as a means of exchange. However, when it comes to the case of dealing with Bitcoin, leverage known as *Margin* and speculation that includes monopolism are both forbidden and Haram unanimously. In the case of other cryptocurrencies that exist on trading platforms, their founder and type of project should be investigated. If the project involves gambling or belongs to a platform that gains profit from leverage such as "BNB"[5], then it is Haram and forbidden in Islam. Hence, the fatwa on cryptocurrencies can't be drawn as a single fatwa of Halal or Haram but depends on the background investigation of the cryptocurrency.

Abu-Bakar (2018) agrees that Bitcoin, among other cryptocurrencies, is permissible due to the prevailing market price on global exchanges of Bitcoin and being accepted for payment at a wide variety of merchants. According to the maxim "The original rule on financial transactions is permissibility" (الأصل في المعاملات الإباحة), it is denoted that financial transactions are permissible unless found in contradiction with the Sharia principles and any form of payment can be considered as money if it is treated as a valuable thing among people and accepted as a medium of exchange.

---

[5] Binance Coin (BNB), a sub-project Cryptocurrency that belongs to Binance, a Cryptocurrency exchange platform. Source: https://coinmarketcap.com/

Referring to Abojeib (2021) in his presented research to the cryptocurrency seminar, cryptocurrencies are assets that take many forms differing in their design and architecture but share their usage of blockchain and cryptography, such that each form affects the jurisprudence ruling on it. Abojeib (2021) stated that to reach a ruling on some cryptocurrency, its software infrastructure must be reliable and secure. It is necessary to check on the background project of cryptocurrency and its financial data to ensure its avoidance of forbidden activities in Islam. The author added that one must investigate the cryptocurrency trading platform used and confirm that the possession and circulation taking place on the platform must not violate Sharia. Lastly, the author recommended forming a committee to follow the recent updates on cryptocurrencies and uniting to conclude a jurisprudence ruling on each form of the cryptocurrencies.

*Opponents*

The Grand Mufti of Egypt, the Turkish government, the Fatwa Center of Palestine, and the UK-based scholar Shaykh Haitham among others, have concluded that Bitcoin and all other cryptocurrencies are Haram and forbidden in Islam (Abu-Bakar, 2018; Hooper, 2017; AboutIslam, 2018). The reasons for disallowing cryptocurrencies can be outlined as follows:

1. Cryptocurrencies have no intrinsic value and no redeemable weight that meets the criteria of Mal in Islam; therefore, they cannot be considered as commodity money.

2.  Cryptocurrency values are unstable and are open to speculation due to the massive rise and drop in prices, including the possibility that they can be used for illegal activities such as money laundering.

3.  Investing in cryptocurrencies is like gambling since it's not guaranteed that miners are going to be successful in solving the mathematical computation to create Bitcoin.

*Neutral*

Islamic scholar AlKhathlan (2021) has recently stated that cryptocurrency needs more examination among Islamic scholars. He believes that modern Islamic scholars have different opinions on cryptocurrencies as they hold uncertainty in some of their aspects and are not transparent. Cryptocurrencies might raise Gharar and risk due to the sudden rises and falls in their price, including their unknown future. Hence, he still hasn't issued a jurisprudence ruling on cryptocurrencies.

Moreover, Abdeldayem et al., (2020) conducted research to investigate cryptocurrency in Islamic finance, particularly in the Gulf Cooperation Council (GCC), and addressed important questions about the Islamic finance view on cryptocurrency and whether cryptocurrency can be considered legal money and a medium of exchange in Sharia. The findings of this study indicate that there is no conclusive answer to the ruling on cryptocurrency in Sharia, therefore the matter is suspended for further research in all its aspects. Furthermore, the authors concluded that cryptocurrencies should be subject to financial and Sharia compliance and oversight so that Muslims worldwide can adopt them, especially in the Gulf countries.

**Cryptocurrency Credibility**

The rapid growth and innovation in the cryptocurrency industry need to be examined due to the potential for fraud from actors manipulating the prices in the marketplace and creating new coins promising misleading benefits. A study by Gandal et al. (2021) investigated cryptocurrency coin and token dynamics, which concluded that most coins and tokens are not steadily traded as they are more likely to be abandoned. They found that 44% of coins were abandoned and 71% resurrected, leaving 18% of coins to fail permanently. Furthermore, they discovered that tokens are less likely to be abandoned because they are a utility to a platform with great economic activity, while coins are speculative instruments.

Several authors have recognized the rise in cryptocurrency scams resulting in immersive financial loss for individuals. For instance, Gao et al. (2020) explored counterfeit cryptocurrencies and measured their impact by analyzing tokens based on the Ethereum blockchain since it does not enforce any restrictions on the names of newly created cryptocurrencies endorsing attackers to use the same identifier as the official cryptocurrency. They have identified 2177 counterfeit tokens that target popular cryptocurrencies, i.e., Bitcoin and Ethereum. They observed that over 7104 victims were deceived by these counterfeit cryptocurrency scams, with an overall financial loss of 17 million dollars. Correspondingly, Xia et al. (2020) further observed cryptocurrency exchange scams through trading platforms such as scam domains and fake apps. They have revealed that most scam domains and fake apps were created and controlled by a small group of attackers, and over 182 blockchain addresses are related to such scams that impacted financial loss of at least 520k dollars.

Finally, A recent study by Bartoletti et al. (2021) extensively reviewed the scientific literature on cryptocurrency scams. They distinguished between two types of scams: *address-reported scams,* which are Bitcoin addresses reported by victims, and *URL-reported scams* from fraudulent websites. They have developed an open-source tool to classify scams and concluded that most scams in their dataset are related to Bitcoin and do not rely on the features of Bitcoin's blockchain but use the original cryptocurrency, i.e. Bitcoin, as a means of payment.

A summary of reviewed points is presented in Table 1. We would like to declare that this section represents the authors' views and opinions and is not considered an official jurisprudence ruling on cryptocurrencies. Due to the limitation of research in our investigation and target domain, we have tackled other lines of work that might help in the direction of our research.

**Table 1**

*Summary of reviewed sources*

| Authors | Purpose | Type of Source | Summary Points |
|---|---|---|---|
| Alsulmi (2020) | Conduct a ruling on cryptocurrencies | Video | • Cryptocurrencies are a commodity and a medium of exchange.<br>• The founder and project of cryptocurrency must be investigated to be free from gambling and leverage.<br>• The ruling on cryptocurrencies depends on the background analysis of the project. |
| Abu-Bakar (2018) | Discuss cryptocurrency and blockchain in Islamic finance | Research | • Any form of payment can be considered money if it is treated as a valuable thing among people and accepted as a medium of exchange.<br>• All financial transactions are permissible if not found to contradict Sharia. |
| Abojeib (2021) | Provide a technical and jurisprudence analysis of cryptocurrencies | Seminar Paper | • Each cryptocurrency takes a different form in its design and architecture but share the use of cryptography and blockchain.<br>• The software infrastructure must be reliable and secure for the cryptocurrency to be permissible.<br>• The cryptocurrency's financial data and trading platform must ensure the permissibility of its possession and circulation in Sharia. |
| Abu-Bakar (2018), Hooper (2017), and | Review Islamic scholars' opponent opinion on cryptocurrencies | Research & website article | • Cryptocurrencies have no intrinsic value and no redeemable weight that meets the criteria of Mal in Islam. |

| AboutIslam (2018) | | | • Cryptocurrencies' values are unstable and are open to speculation due to the massive rise and drop in prices.<br>• The possibility of cryptocurrencies being used for illegal activities such as money laundering.<br>• They think that investing in cryptocurrencies is like gambling due to the uncertainty of miners' success. |
|---|---|---|---|
| AlKhathlan (2021) | Conduct a ruling on cryptocurrencies | Video | • Cryptocurrencies hold uncertainty and the ruling is not transparent.<br>• Cryptocurrencies raise Gharar and risk due to the rapid rise and drop of their prices and their unknown future of them. |
| Abdeldayem et al. (2020) | Investigate cryptocurrency in Islamic finance, particularly within the Gulf Cooperation Council (GCC) | Research | • Addressed questions regarding the Islamic finance view on cryptocurrency and whether they are considered legal money and a medium of exchange in Sharia.<br>• The results indicate no conclusive answer to the ruling on cryptocurrency, and the matter is suspended for further research.<br>• Cryptocurrencies should be subject to financial and Sharia compliance so that Muslims around the world can adopt them. |
| Gandal et al. (2021) | Analyze the dynamics of coins and tokens in the cryptocurrency industry | Research | • 44% of cryptocurrency coins were abandoned, and 71% resurrected, leaving 18% of coins to fail permanently.<br>• Tokens are less frequent to be abandoned due to them being represented as a utility, while coins are speculative instruments. |
| Gao et al. (2020) | Track and identify fraudulent behaviors related to counterfeit cryptocurrencies | Research | • They have identified 2177 tokens that target popular cryptocurrencies such as Bitcoin and Ethereum.<br>• Counterfeit cryptocurrencies deceived over 7104 victims with 17 million dollars of an overall financial loss. |
| Xia et al. (2020) | Identify and characterize cryptocurrency exchange scams | Research | • Exchange scams take place on trading platforms and are in the form of fake apps and scam domains.<br>• A small number of attackers controls most fake apps and scam domains.<br>• Over 182 blockchain addresses are related to scams with at least 520k dollars of financial loss. |
| Bartoletti et al. (2021) | Perform an extensive review of scientific literature on cryptocurrency scams | Research | • Distinguished between two types of scams: *address-reported scams* which are Bitcoin addresses reported by victims, and *URL-reported scams* from fraudulent websites.<br>• Most scams are related to Bitcoin and do not rely on the features of the blockchain, but the original cryptocurrency as a means of payment. |

## Methodology

CryptoHalal system is built using a trained ML model that classifies cryptocurrencies entered by users into Halal or Haram with a supported explanation based on selected features. The block diagram in Figure 5 presents the architecture of the CryptoHalal system, including the subsystems and users. In this architecture, users interact with the system through an interface to access separate services for each user. User services use the trained CryptoHalal model to classify the entered cryptocurrency by users, also displays the retrieved data from the database. Islamic scholars' services revolve around the access and display of the database.

Once the user enters the cryptocurrency's name, the system checks for a match in the local database, and if a match exists, the output fetched from the database will be displayed to the user. Otherwise, the system activates CoinMarketCap API using the "Metadata" endpoint, which returns the cryptocurrency information to retrieve the official website URL and scrape the website using Beautifulsoup[6]. Then, we will apply text pre-processing steps which include: removing all HTML tags, tokenization, removing stop words, and stemming. This will lead to a raw text file ready for feature extraction. Finally, the cryptocurrency will be classified by our model and an output will be presented to the user as "Probably Halal" or "Probably Haram" with respect to the classification.

---

[6] a library in Python to fetch data from websites into a raw HTML file

**Figure 5**

*CryptoHalal system architecture diagram*

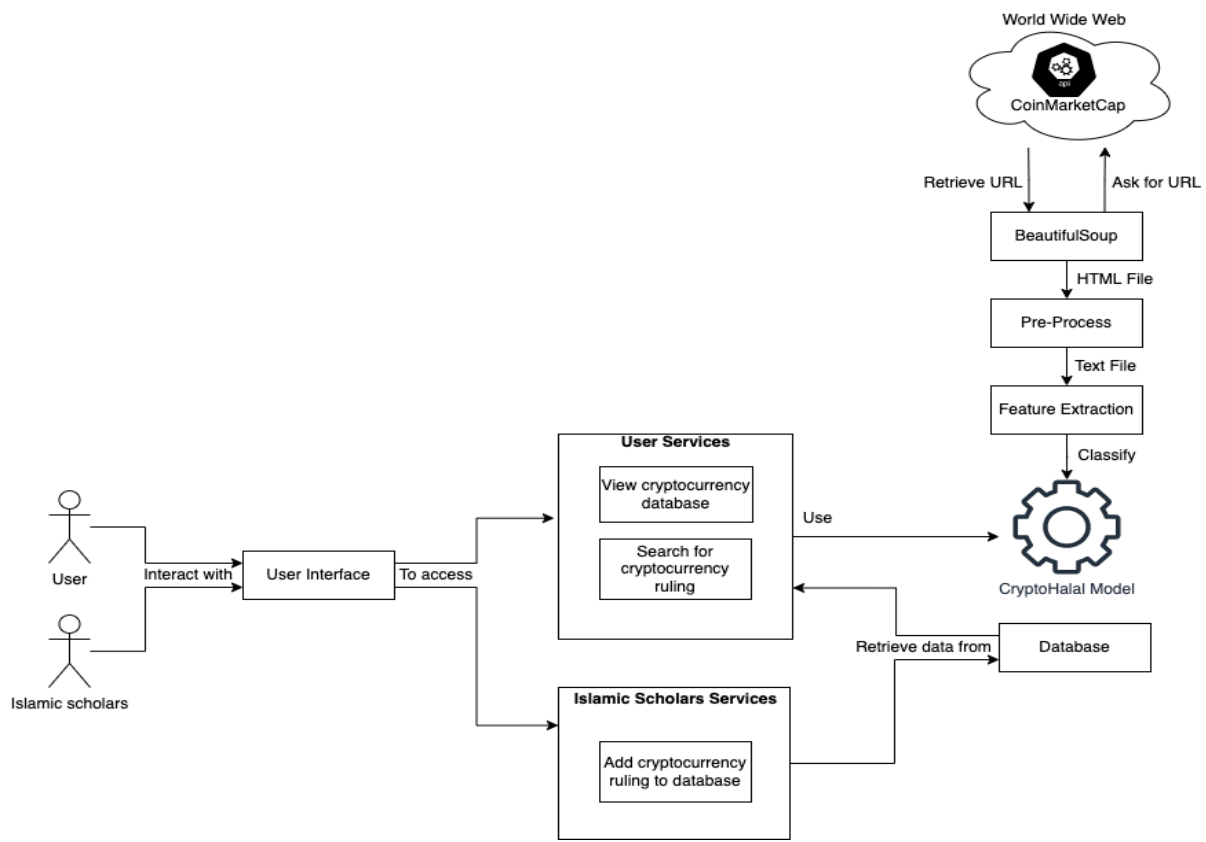

## CryptoHalal Users

CryptoHalal system is used by two types of users: end-users and Islamic scholars. End-users can view the database which contains all registered cryptocurrencies with their ruling and explanation, and search for a cryptocurrency ruling by entering its name or ticker[7]. Figure 6 illustrates the user interfaces designed for the system. End-users view (A) which consists of a search box and some additional services such as viewing the database, while (C) and (D) demonstrate a sample run of the presented output to the user that was fetched from cryptocurrency features. Islamic scholars can access the page viewed in Figure 6 (B) through a log-in page, where they can assist in entering cryptocurrency rulings into the database, as well as modifying cryptocurrency rulings.

---

[7] A short combination of letters used to represent cryptocurrency, e.g. BTC is Bitcoin's ticker.

**Figure 6**

*User interfaces for CryptoHalal website, (A) is End-user interface, (B) Islamic Scholar interface, (C) is cryptocurrency Halal output, and (D) is cryptocurrency Haram output.*

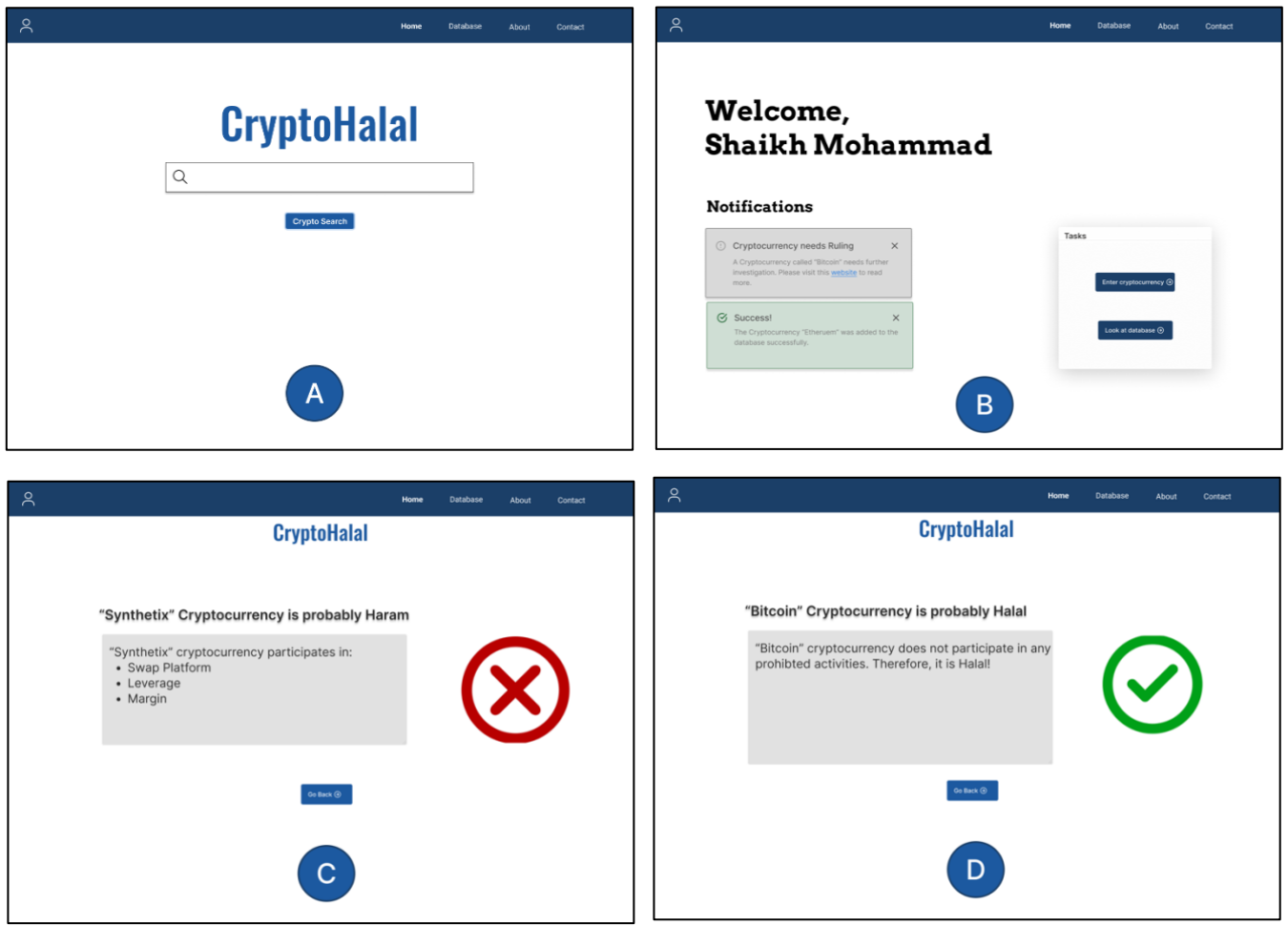

## Data Collection

To date, there are no published datasets that contain the classification of cryptocurrencies as Halal or Haram. Our data was collected from a Telegram channel called CRYPTOHALAL (كريبتو حلال)[8] supervised by Islamic scholars, which is specialized in the observation and legal audit of the cryptocurrency market to assist Muslims with the jurisprudence ruling on cryptocurrencies. Figure 7 shows an example of cryptocurrency ruling cards published by the channel, which contains the description, services, and uses of the cryptocurrency. Our dataset consists of 106 cryptocurrencies classified into 56 Halal and 50 Haram cryptocurrencies. The dataset contains 20 features that take binary values (1 as yes and 0 as no) for each cryptocurrency. A full explanation of features can be seen in Table 2. We would like to add that the dataset has been downloaded from GitHub[9].

---

[8] https://t.me/cr_halal [Date accessed: 19/4/2022]

**Figure 7**

*Example of cryptocurrency ruling cards [Source: CRYPTOHALAL Telegram channel]*

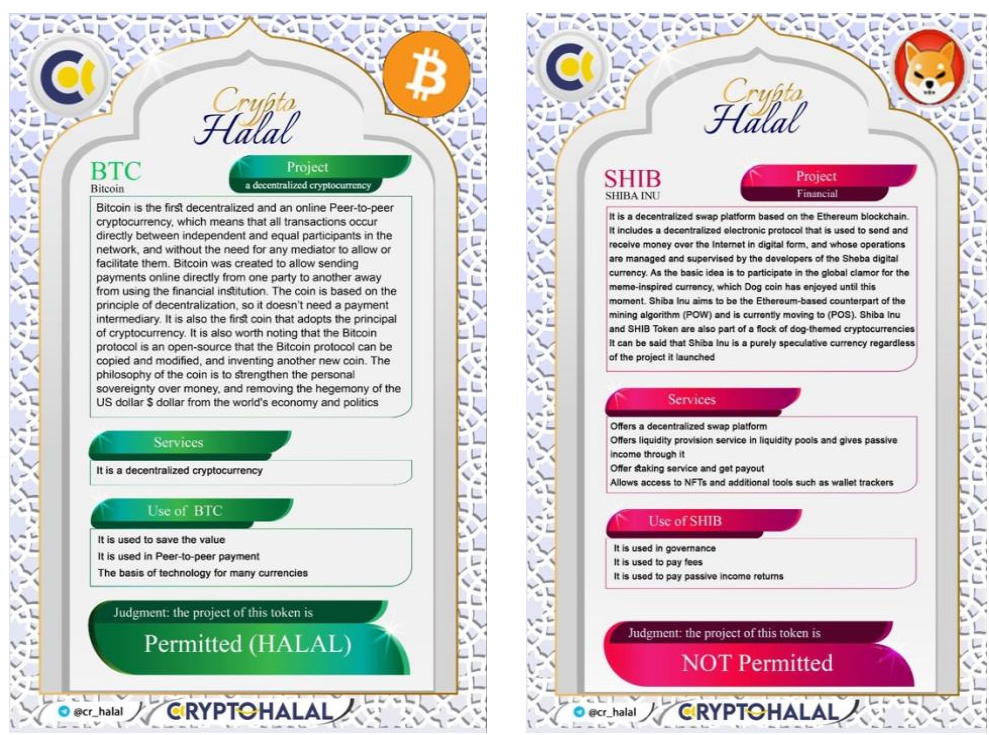

---

**Feature Selection and Engineering**

In this work, we used 20 handcrafted features derived from cryptocurrency ruling cards. The features were selected according to a manual analysis of cryptocurrency rulings, such that features were extracted based on keywords recurrence in the collected data. All features are numeric besides the "Ruling" feature as the class label. Table 2 shows the list of extracted features that were used to train the classifiers along with their descriptions.

When analyzing the dataset, we have found that 45 out of 50 Haram cryptocurrencies' projects use Decentralized Finance (DeFi) and liquidity pools, whereas none of the Halal cryptocurrencies participate in lending & borrowing services, leverage, margin, and prediction market. Interestingly, none of the Haram cryptocurrencies' projects is technical as the majority of the Haram cryptocurrencies' projects are purely financial. Our assumption is that financial cryptocurrency projects may involve Riba and gambling to increase interest. Therefore, based on the analysis we have separated features based on their priorities that impact the ruling of cryptocurrencies (See Table 3). High Priority indicates features that have only Haram cryptocurrencies, and Low Priority indicates features consisting of both Halal and Haram cryptocurrencies. This separation will assist in determining the cryptocurrency's degree of compliance with Sharia. For example, if the majority of the high-priority features in the cryptocurrency are satisfied, then it is more likely to be Haram with an outcome of "Probably Haram", otherwise it is more likely to be Halal with an outcome of "Probably Halal".

**Table 2**

*List of features used for classification*

| Features | Explanation |
|---|---|
| Coin | Cryptocurrency ticker name |
| Ruling | Indicates if a cryptocurrency is Halal or Haram |
| PoW | Cryptocurrency uses proof-of-work protocol |
| Ethereum_Blockchain | Cryptocurrency is based on Ethereum blockchain network |
| PoS | Cryptocurrency uses proof-of-stake protocol |
| DeFi | Cryptocurrency project uses DeFi |
| Speculation | Cryptocurrency project is based on speculation |
| Staking | Cryptocurrency offer staking services |
| Swap_Platform | Cryptocurrency offers a decentralized swap platform to swap at best price |
| Liquidity | Cryptocurrency contains liquidity pools |
| Lending | Cryptocurrency provides lending services |
| Borrowing | Cryptocurrency provides borrowing services |
| Prediction_Market | Cryptocurrency participates in the prediction market based on bets |
| Leverage | Cryptocurrency project designed for gaining leverage |
| Margin | Cryptocurrency project designed for margin trading |
| Yield_farming | Cryptocurrency provides yield farming services with passive income |
| Governance | Cryptocurrency offers governance of the protocol for the future of the cryptocurrency |
| Financial_Project | Cryptocurrency project is pure financial without additional project services |
| Technical_Project | Cryptocurrency project is technical that offers on-chain Decentralized App services and software development tools |
| Service_Project | Cryptocurrency project is based on services such as betting and media |

**Table 3**

*Features' priorities*

| High Priority | Low priority |
|---|---|
| Speculation | Staking |
| Borrowing | Swap_Platform |
| Prediction | Liquidity |
| Leverage | Lending |
| Margin | Governance |
| Yield_Farming | Financial_Project |
| | Services_Project |
| | DeFi |

**CryptoHalal Model**

We experimented with three supervised ML models, compared the performance, and chose the best classifier to be used in the CryptoHalal system. The models are: Support Vector Machine (SVM) (Mtetwa et al., 2017), Naïve Bayes, and Logistic Regression (Luo, 2021). These three classifiers are well-suited for classification problems with small datasets and perform exquisitely such as in our case.

A previous work by Mtetwa et al., (2017) used an SVM model for classifying web pages as fraudulent or not. They extracted their data using web crawlers and used a feature list created by (Abbasi et al., 2010) associated with fake and fraudulent websites. They have decreased the number of features used in their evaluation and have come to the conclusion that using all features in their feature list can increase the accuracy of the model but may lead to a decrease in computational efficiency.

Another work by Luo (2021) used SVM, NB, and LR for classifying documents into categories. However, based on the author's experiment SVM classifier outperformed the rest of the techniques based on the evaluated measures. In addition, Luo used WEKA experimental tool, similar to our case. Therefore, based on the presented previous works we chose the ML models that best suited our experiment and data.

**Evaluation**

We used WEKA (Waikato Environment for Knowledge Analysis)[10] to train and classify our data. Table 3 presents detailed classifiers' performance on the training set using 10-fold cross-validation for the three ML models. The evaluation measures used are: Accuracy, Precision, Recall, and F-Measure.

---

[10] Gautam, Shivani, Chetan Sharma, and Vinay Kukreja. "Handwritten Mathematical Symbols Classification Using WEKA." In Applications of Artificial Intelligence and Machine Learning, pp. 33-41. Springer, Singapore, 2021.

In terms of accuracy, we can see that SVM and NB got the same rate whereas LR performed poorly compared to them due to a higher amount of incorrectly classified instances. The only difference between SVM and NB performances is the precision with a slight fractional difference of 0.1 in favor of SVM.

When examining the output predictions, we found some incorrectly classified cryptocurrencies in SVM and NB. Namely, GNO and AUDIO. For GNO coin, we couldn't find a supported explanation on why the models predicted it as Halal, even though the coin participated in prediction market services which is similar to gambling. As for AUDIO, we found out that it does not contain the features that are famously involved in Haram cryptocurrencies, but the coin's project is involved in the music industry, thus it is Haram. LR model had only one incorrectly classified cryptocurrency in common with SVM and NB, which is AUDIO. Therefore, we chose the SVM model for the CryptoHalal system's classifier due to its acceptable performance based on the used evaluation measures.

**Table 4**

*Performance of classifiers*

| Parameter | SVM | NB | LR |
|---|---|---|---|
| Accuracy (%) | 97.17 % | 97.17 % | 94.34 % |
| Precision (%) | **97.3 %** | 97.2 % | 94.3 % |
| Recall (%) | 97.2 % | 97.2 % | 94.3 % |
| F-Measure (%) | 97.2 % | 97.2 % | 94.3 % |

**Conclusion**

Cryptocurrency jurisprudence ruling classification is an essential task in the field of Islamic finance. The application of the CryptoHalal system can help Muslims deal with, buy, and trade cryptocurrencies in a Sharia-compliant way. In this research, we presented an intelligent system that will serve Muslim cryptocurrency investors and traders worldwide. This may be the first research that defines and investigates cryptocurrency features to conclude a jurisprudence ruling on them. In addition to designing a system that serves two types of users (end-users and scholars).

In the CryptoHalal system, we applied three ML models to compare performances and have concluded that SVM and NB gave us similar performances, whereas LR performed poorly compared to them. Therefore, we chose SVM as our system's classifier, mainly because we believe that it will outperform other classifiers with the expansion of our dataset.

While our system brings convenience to Muslims worldwide, it also brings huge risks threatening the security of the system from being illegally exploited by unauthorized personnel. Thus, it is important to shed the light on the security performance of the system to avoid cyber-attacks and improve the efficiency of vulnerability recovery and management. Additionally, cryptocurrencies haven't been passed regulations in Saudi Arabia, despite having them passed in many countries, including the GCC countries i.e. Bahrain and the United Arab Emirates (Abdeldayem & Aldulaimi 2020), and that plays a vital role in people's acceptance of investing in cryptocurrencies.

The limitation of this work entails the small dataset and the manual extraction of features. Thus, we aim to overcome these limitations by increasing the dataset, applying Deep Learning models for feature extraction, and applying advanced feature engineering techniques such as Information Gain and Term Frequency-Inverse Document Frequency

(TF-IDF).